\documentclass[12pt,preprint]{emulateapj}
\usepackage{epsfig}

\def\gsim{\;\rlap{\lower 2.5pt
 \hbox{$\sim$}}\raise 1.5pt\hbox{$>$}\;}

\def\lsim{\;\rlap{\lower 2.5pt
   \hbox{$\sim$}}\raise 1.5pt\hbox{$<$}\;}

\def\dcI{\delta_{cd}}
\def\dcII{\delta_{cp}}
\def\dn{\delta_n}

\def\dN{\delta_N}
\def\de{\delta_e}
\def\dcIe{\dcI-\de}
\def\dcIn{\dcI-\dn}
\def\dcIIN{\dcII-\dN}
\def\dcIIn{\dcII-\dn}
\def\d7{\delta_{7-FOF}}

\def\Sji{S_j-S_i}
\def\Sjm{S_j-S_m}
\def\Smj{S_m-S_j}
\def\Sjn{S_j-S_n}
\def\Snj{S_n-S_j}
\def\Sim{S_i-S_m}
\def\Smi{S_m-S_i}
\def\Sin{S_i-S_n}
\def\Snm{S_n-S_m}
\def\Sde{S_d-S_e}
\def\SNn{S_N-S_n}
\def\Spd{S_p-S_d}
\def\SNj{S_N-S_j}

\def\Sni{S_n-S_i}

\def\Spe{S_p-S_e}
\def\d{\delta}
\def\D{\Delta}
\def\eps{\epsilon}

\newcommand{\Deltafunc}[2]{\Delta(S_{#1},S_{#2})}

\newcommand{\explarge}[1]{\exp\left[#1\right]}
\newcommand{\erfc}[1]{{\mathrm {erfc}}\left[#1\right]}

\newcommand{\II}[5]{\Pi_{#1}^{#2}(#3;#4;#5)}
\newcommand{\IIzero}[4]{\II{0}{#1}{#2}{#3}{#4}}
\newcommand{\IIzeroI}[3]{\IIzero{\dcI}{#1}{#2}{#3}}
\newcommand{\IIzeroII}[3]{\IIzero{\dcII}{#1}{#2}{#3}}

\newcommand{\IIep}[4]{\II{\epsilon}{#1}{#2}{#3}{#4}}
\newcommand{\IIepI}[3]{\IIep{\dcI}{#1}{#2}{#3}}
\newcommand{\IIepII}[3]{\IIep{\dcII}{#1}{#2}{#3}}

\newcommand{\intn}[2]{\int_{-\infty}^{#1}d#2}
\newcommand{\intnI}[1]{\intn{\dcI}{#1}}
\newcommand{\intnII}[1]{\intn{\dcII}{#1}}

\newcommand{\intS}[3]{\int_{#1}^{#2}d#3}
\newcommand{\partialone}[1]{\frac{\partial}{\partial #1}}
\newcommand{\partialtwo}[2]{\frac{\partial^2}{\partial #1\partial #2}}
\newcommand{\partialthree}[3]{\frac{\partial^3}{\partial #1\partial #2\partial #3}}

\newcommand{\ratioI}[2]{\frac{#1}{(#2)^{3/2}}}
\newcommand{\ratioII}[2]{\frac{(#1)^2}{2(#2)}}

\def\Eone{{\mathbf {E_1}}}
\def\Etwo{{\mathbf {E_2}}}
\def\Ethree{{\mathbf {E_3}}}
\def\Erone{{\mathbf {Er_1}}}
\def\Ertwo{{\mathbf {Er_2}}}
\def\Erthree{{\mathbf {Er_3}}}
\def\IIone{{\mathbf {\Pi}}}
\def\Fone{{\mathbf {F_1}}}
\def\Ftwo{{\mathbf {F_2}}}

\def\gsim{\;\rlap{\lower 2.5pt \hbox{$\sim$}}\raise 1.5pt\hbox{$>$}\;} 
\def\lsim{\;\rlap{\lower 2.5pt \hbox{$\sim$}}\raise 1.5pt\hbox{$<$}\;} 
\def\ie{{\it i.e.}}

\shorttitle{Environmental Dependence in Excursion Set Model}
\shortauthors{Zhang, Ma, Riotto}

\begin{document}
\title{Dark Matter Halo Assembly Bias: Environmental Dependence in the Non-Markovian Excursion Set Theory}

\author{Jun Zhang}
\affil{Center for Astronomy and Astrophysics, Department of Physics and Astronomy, Shanghai Jiao Tong University, 955 Jianchuan road, Shanghai, 200240, China} 
\email{betajzhang@sjtu.edu.cn}

\author{Chung-Pei Ma}
\affil{Department of Astronomy, University of California, Berkeley, CA 94720, USA}

\author{Antonio Riotto}
\affil{Department of Theoretical Physics and Center for Astroparticle Physics (CAP),  24 quai E. Ansermet, CH-1211 Geneva, Switzerland}

\begin{abstract}

  In the standard excursion set model for the growth of structure, the
  statistical properties of halos are governed by the halo mass and are
  independent of the larger scale environment in which the halos reside.
  Numerical simulations, however, have found the spatial distributions of
  halos to depend not only on their mass but also on the details of their
  assembly history and environment.  Here we present a theoretical
  framework for incorporating this ``assembly bias'' into the excursion set
  model.  Our derivations are based on modifications of the path integral
  approach of Maggiore \& Riotto (2010) that models halo formation as a
  non-Markovian random walk process.  The perturbed density field is
  assumed to evolve stochastically with the smoothing scale and exhibits
  correlated walks in the presence of a density barrier.  We write down
  conditional probabilities for multiple barrier crossings, and derive from
  them analytic expressions for descendant and progenitor halo mass
  functions and halo merger rates as a function of both halo mass and the
  linear overdensity $\delta_e$ of the larger-scale environment of the
  halo.  Our results predict a higher halo merger rate and higher
  progenitor halo mass function in regions of higher overdensity,
  consistent with the behavior seen in $N$-body simulations.
 
\end{abstract}


\section{Introduction}
\label{intro}

In hierarchical cosmological models such as $\Lambda$CDM, dark matter halos
of lower mass form earlier on average than more massive halos.  The virial
mass of halos is a key parameter that governs many properties of galaxies
and their host halos, e.g., galaxy morphology and color, baryonic feedback
processes, formation redshift, and halo occupation number.  Recent
numerical simulations, however, have shown that a halo's local environment
-- in addition to its mass -- also affects the formation processes.  At a
{\it fixed} mass, older halos are found to cluster more strongly than more
recently formed halos \citep{Gottlober01, ST04, Gao05, Harker06,
  Wechsler06, Jing07, Wang07, GW07, Maulbetsch07, Angulo08,
  dalal08,li08}. Other halo properties such as concentration, spin, shape,
velocity structure, substructure mass function, merger rates, and halo
occupation distribution have also been shown to vary with halo environment
(e.g., \citealt{Avila05, Wechsler06, Jing07, GW07, Bett07, Wetzel07, FM09,
  FM10, FW10, zentner13}).

In comparison, the formation and properties of dark matter halos depend
only on the mass and not environment in the extended Press-Schechter and
excursion set models \citep{PS74, Bond91, LC93}.  These models are used
widely for making theoretical predictions of halo and galaxy statistics and
for Monte Carlo constructions of merger trees.  The lack of environmental
correlation arises from the Markovian nature of the random walks in the
excursion set model: the change of the matter over-density as a function of
the smoothing scale is treated as a Markovian process, which by definition
decouples the density fluctuations on small (halo) and large (environment)
scales.  This limitation stems from the use of the Fourier-space tophat
window function as the mass filter.  When a Gaussian window function is
used, for instance, \citet{Zentner07} finds an environmental dependence in
the halo formation redshift, but the dependence is {\it opposite} to that
seen in the numerical simulations cited above. Several other attempts at
incorporating environmental effects into the excursion set model were not
able to reproduce the correlations seen in the simulations (e.g.,
\citealt{Sandvik07, DesJacques08}).

In this paper we aim to derive analytic expressions for halo statistics
that depend on halo mass as well as its large-scale environmental density.
To achieve this goal, we begin with the non-Markovian extension of the
excursion set model by \cite{MR1} (MR10 hereafter).  In this approach, a
path integral formalism is used to perform perturbative calculations for
non-Markovian processes of Gaussian fields. A key quantity is the
probability that the smoothed matter over-density remains below a critical
value down to a certain mass scale (equation~[40] of MR10). They show that
this quantity can be written as a multi-variable integral of a Gaussian
distribution function, which can be worked out exactly in the Markovian
case, and perturbatively for weakly non-Markovian processes (see \S 3, 4, 5
of MR10 for details).  This probability can be used to derive the
first-crossing rate for the halo mass function as shown in equation~(42) of
MR10.

To introduce environmental dependence, we modify equation~(40) of MR10 by
first isolating (\ie, not integrating out) the dependence of the matter
overdensity on the specified environmental scale in this equation.  We then
add to the path integral a portion that is between the descendant and
progenitor halo mass scales with a slightly higher critical value for halo
identification (corresponding to the halo formation criteria at a slightly
higher redshift).  The resulting new probability is a function of the
environmental density and the descendant and progenitor halo masses. Its
derivative with respect to the descendant and the progenitor masses yields
the conditional halo mass function as a function of the overdensity of the
larger-scale environment, which will be the main result of this paper.
In \S\ref{MergerRates}, we provide a summary of the excursion set model and
the path-integral approach to the non-Markovian extension.  In \S~3, we
introduce the formalism and perform the main calculation, including the
simplification of the final result in the limit of the large scale
environment.

\section{Non-Markovian Extension to the Excursion Set Model}
\label{MergerRates}

\subsection{Summary of the excursion set model}

At any given time $t$ and position ${\bf x}$, a virialized dark matter halo
is formed in the excursion set model if the linear mass overdensity
$\delta({\bf x}, R)$ smoothed on the scale of the halo size $R$ exceeds a
threshold $\delta_c$ that is determined by the spherical collapse model,
and if no larger smoothing scales meet the criterion.  The smoothed density
field is given by
\begin{equation}
\d({\bf x},R) =\int d^3x'\,  W(|{\bf x}-{\bf x}'|,R)\, \d({\bf x}')\, ,
\end{equation}
where $\delta({\bf x}) = \rho({\bf x})/\bar\rho - 1$ is the density
contrast about the mean mass density $\bar\rho$ of the universe, $W(|{\bf
  x}-{\bf x}'|,R)$ is the smoothing filter function, and $R$ is the
smoothing scale.  When $W$ is a tophat function in $k$-space, the
over-density traces out the smoothing scale as a Markovian random walk
process.  Instead of $R$, the variance $S$ of the density field is often
used to denote the length (or mass) scale, where
\begin{equation}
   S(R) \equiv \sigma^2(R) = \int \frac{d^3k}{(2\pi)^3}\, P(k) \tilde{W}^2(k, R) \,.
\label{S}
\end{equation}
Here $P(k)$ is the power spectrum of the matter density fluctuations in a
given cosmological model, and $\tilde{W}$ is the Fourier transform of the
filter function $W$.  As the smoothing radius $R$ goes to infinity, $S(R)$
goes to zero.  In hierarchical models of structure formation such as the
$\Lambda$CDM model, $S$ is a monotonically decreasing function of $R$.  The
variables $S$, $R$, and the associated mass, $M=(4/3)\pi R^3 \bar{\rho}$, can
therefore be used interchangeably.

In the standard excursion set model, the first crossing distribution of
random walks with a constant barrier $\delta_c$ determines the halo mass
function.  Further refinement is achieved by the ellipsoidal collapse model
with a scale-dependent $\delta_c$ \citep{Sheth01,ST02}, or a diffusing
barrier \citep{Robertson09, MR2}.  The resulting halo mass functions are
found to agree reasonably well with $N$-body simulation results (e.g.,
\citealt{Tinker08, Ma11}).

In addition to the halo mass function, the excursion set model also
predicts the halo assembly history. As the linear density field grows with
time, halos are identified on increasingly larger mass scales, which
signifies the gain of dark matter mass through mergers or accretion.
Statistics such as the halo merger rates, progenitor mass functions, and
their relations with the large scale environmental density can all be
worked out in this framework.  

The calculation of the halo statistics typically treats the change of the
smoothed linear density field $\delta$ with a decreasing smoothing scale
$S(R)$ as a Markovian process, in which each step of the random walk is
uncorrelated with the previous one.  The Markovian assumption therefore
decouples the linear density fluctuations below and beyond the halo mass
scale, causing the halo properties, such as its formation time and merger
rate, to be independent of the density of the halo environment.  This
assumption greatly simplifies the calculations and has led to a number of
useful analytic results.  The Markovianity of the process, however, relies
on the density smoothing filter being a tophat function in $k$-space, which
does not correspond to a well-defined halo mass in real space.  In
addition, the decoupling between halo mass and halo environment is not seen
in numerical simulations.

\subsection{Introduce non-Markovianity}

A difficulty of the excursion set model is that an unambiguous relation
between the smoothing radius $R$ and the mass $M$ of the corresponding
collapsed halo only exists when the filter is a tophat function in real
space: $M(R)=(4/3)\pi R^3 \bar{\rho}$.  For all other filter
functions (e.g., tophat in $k$-space, Gaussian), it is impossible to
associate a well-defined mass $M(R)$ (see, e.g.,
\citealt{Bond91, Zentner07}).  

To deal with this problem, \cite{MR1} uses a path integral approach to
compute the probability associated with each trajectory $\d(S)$ and sum
over all relevant trajectories.  For convenience, the time variable is
first discretized and the continuum limit is taken at the end.
Specifically, we discretize the interval $[0,S]$ in steps $\D S=\eps$, so
$S_k=k\eps$ with $k=1,\ldots n$, and the end point is $S_n\equiv S$.  A
trajectory is defined by the collection of values $\{\delta_1,\ldots
,\delta_n\}$, such that $\delta(S_k)=\delta_k$.  All trajectories start at
a value $\d_0$ at ``time'' $S=0$.

The basic quantity in this approach is the probability density in the space
of trajectories, defined as
\begin{equation}
\label{define_w}
W(\delta_0; \delta_1, \ldots , \delta_n; S_n)
\equiv \left\langle\delta_D[\delta(S_1)-\delta_1]\ldots\delta_D[\delta(S_n)-\delta_n]\right\rangle
\end{equation}
where $\delta_D$ is the Dirac delta function, and all trajectories start
from $\delta_0$ at $S=0$.  For a Gaussian random density field, 
the only non-zero component in $W$ is the connected two-point correlator 
$\langle\delta_j\delta_k\rangle_c$, and $W$ can be transformed into:
\begin{eqnarray}
\label{W2}
&& W(\delta_0; \delta_1, \ldots , \delta_n; S_n)
= \int_{-\infty}^{\infty}\frac{d\lambda_1}{2\pi}\ldots\frac{d\lambda_n}{2\pi}\\ \nonumber
&& \quad \times \exp\left(i\sum_{j=1}^{n}\lambda_j\delta_j-\frac{1}{2}\sum_{j,k=1}^{n}\lambda_j\lambda_k\langle\delta_j\delta_k\rangle_c\right) \,.
\end{eqnarray}

If the density smoothing filter is a top-hat function in $k$-space, the
evolution of $\delta(S)$ is Markovian, and the density correlation is:
\begin{equation}
\label{density_c}
\langle\delta_i\delta_j\rangle_c={\mathrm{min}}(S_i, S_j) \,.
\end{equation} 
In this case, the integrals in equation~(\ref{W2}) can be
worked out directly to give
\begin{eqnarray}
\label{W3}
&& W^{gm}(\delta_0; \delta_1, \ldots , \delta_n; S_n)\\ \nonumber
&& \quad =\frac{1}{(2\pi\epsilon)^{n/2}}\exp\left[-\frac{1}{2\epsilon}\sum_{i=0}^{n-1}(\delta_{i+1}-\delta_i)^2\right] \,,
\end{eqnarray}
where the superscript ``gm'' refers to the ``Gaussian and Markovian''
case. When the density smoothing filter is {\it not} a top-hat function in
$k$-space, e.g., a top-hat function in real space or a Gaussian function,
MR10 showed that an additional term appeared in the density correlation:
\begin{equation}
\label{Delta_ij}
\langle\delta_i\delta_j\rangle_c={\mathrm{min}}(S_i, S_j)+\Delta(S_i,S_j)  \,,
\end{equation} 
where $\Delta(S_i,S_j)$ is well approximated by
\begin{eqnarray}
\label{Delta_define}
&&\Delta(S_i, S_j)\approx \kappa \, \frac{S_{min}(S_{max}-S_{min})}{S_{max}}, \\ \nonumber
&&S_{max}={\mathrm{max}}(S_i, S_j), \quad S_{min}={\mathrm{min}}(S_i, S_j) \,.
\end{eqnarray}
The parameter $\kappa$ characterizes the non-Markovian process, whose value
depends on the shape of the smoothing filter, e.g., $\kappa \simeq 0.44$
for a top-hat function in real space, and $\kappa \approx 0.35$ for a
Gaussian function.

For convenience, we use $\Delta_{ij}$ to denote $\Delta(S_i, S_j)$ in
this paper.  In the non-Markovian case, equation~(\ref{W2}) can
be expanded perturbatively into
\begin{eqnarray}
\label{W4}
&&W(\delta_0; \delta_1, \ldots , \delta_n; S_n)\\ \nonumber
&& \quad \approx \left(1+\frac{1}{2}\sum_{i,j=1}^{n}\Delta_{ij}\frac{\partial^2}{\partial\delta_i\partial\delta_j}\right)
W^{gm}(\delta_0; \delta_1, \ldots , \delta_n; S_n) \,.
\end{eqnarray}
We will use equation~(\ref{W4}) for the rest of paper, keeping in mind that
this relation only includes the leading order non-Markovian corrections.

\section{Main Derivation}
\label{formalism}

\subsection{Introduce the environmental variable}
\label{Define_rate}

The new ingredient that we will introduce into the non-Markovian excursion
set model is the linear overdensity, $\de$, that quantifies the larger-scale
environment of a dark matter halo.  We denote the smoothing scale over which
$\de$ is evaluated as $S_e$, where $S$ is defined in equation~(\ref{S}).

Throughout the paper, we use subscripts ``e'', ``d'', and ``p'' to denote
environment, descendant, and progenitor, respectively.  We consider a
descendant halo of mass $M_d$, or $S_d=S(M_d)$, that formed at redshift
$z_d$ when the barrier height is $\dcI=\delta_c/D(z_d)$, where
$\delta_c=1.68$ and $D(z)$ is the linear growth function.  We consider the
probability for the descendant halo to have a progenitor halo of mass
$M_p$, or $S_p=S(M_p)$, that formed at a higher redshift $z_p$ when the
barrier height is higher: $\dcII=\delta_c/D(z_p)$.  We adopt the convention
that the critical overdensity, instead of the linear overdensity, is a
function of redshift. The linear overdensity is always evaluated at
redshift zero, including that on the environmental scale.

As an initial setup, we define three events A, B, and C as follows.\\

\noindent {\bf A:} At a location of interest, the overdensity smoothed over a scale $S_e$ (centered on the location) is $\de$.\\

\noindent {\bf B:} At the same location as in A, a halo of mass $S_d$ forms at redshift $z_d$, corresponding to barrier $\dcI=\delta_c/D(z_d)$, where $S_d > S_e$.\\

\noindent {\bf C:} At the same location as in A, a progenitor halo of mass $S_p$ forms at redshift $z_p$, corresponding to barrier $\dcII=\delta_c/D(z_p)$, where $z_p>z_d$ and $S_p > S_d > S_e$.\\

We then define the following probabilities that relate the three events above:\\

\noindent 1. $P(A)d\de$ is the probability that the linear overdensity smoothed over
scale $S_e$ is between $\de$ and $\de+d\de$.  For a Gaussian field, we have
the simple relation
\begin{equation}
  P(A)= \exp{(-\delta_e^2/2S_e)}/\sqrt{2\pi S_e} \,.
\end{equation}

\noindent 2. $P(A, B)d\de dS_d$ is the probability that a halo of mass between $S_d$
and $S_d+dS_d$ forms at redshift $z_d$, and at the halo location, the
linear overdensity smoothed over a larger scale $S_e$ is between $\de$ and $\de+d\de$. 
More explicitly, we have
\begin{equation}
P(A, B) = P_{AB}(S_d, z_d, S_e, \de) \,.
\end{equation}


\noindent 3. $P(A, B, C)d\de dS_ddS_p$ is the probability that a halo of mass between
$S_d$ and $S_d+dS_d$ forms at redshift $z_d$, and the mass of this halo at
an earlier redshift $z_p$ is in progenitor of mass between $S_p$ and $S_p+dS_p$,
and at the halo location, the linear overdensity on scale of $S_e$ is between
$\de$ and $\de+d\de$.  More explicitly, we have
\begin{equation}
P(A, B, C) = P_{ABC}(S_p, z_p, S_d, z_d, S_e, \de) \,.
\end{equation}

Our goal is to derive expressions for the following conditional
probabilities that depend on the halo environment parameterized by $\delta_e$ and $S_e$: \\

\noindent 1. $P(B|A)dS_d$ is the probability that a halo of mass between $S_d$ and
$S_d+dS_d$ forms at redshift $z_d$ in an environment of linear overdensity
$\de$ on scale of $S_e$.  More explicitly, we have
\begin{equation}
P(B|A) = P_{(B|A)}\left(S_d, z_d\vert S_e, \de\right) \,.
\end{equation}
As we show in Sec.~3.2, this quantity is simply related to the
environment-dependent halo mass function. \\

\noindent 2. For a halo of mass $S_d$ forming at redshift $z_d$, located in the
center of an environment of scale $S_e$ and linear overdensity $\de$,
$P(C\vert A, B)dS_p$ is the probability that the mass of this halo at an
earlier redshift $z_p$ is in progenitor of mass between $S_p$ and
$S_p+dS_p$.  More explicitly, we have
\begin{equation}
 P(C\vert A, B)=P_{(C|A, B)}\left(S_p, z_p\vert S_d, z_d, S_e, \de\right) \,.
\end{equation}
As we show in Sec.~3.2, this quantity is simply
related to the environment-dependent progenitor mass function and the halo merger rate.
\\

The probabilities are related by $P(B|A)=P(A,B)/P(A)$, $P(C|A, B)=P(A,
B, C)/P(A, B)$.  When the smoothing filter is chosen to be a top-hat
function in $k$-space, the random walk is a Markovian process.
The environmental dependence drops out in this case, and
we have $P(C|A,B)=P(C|B)$, which is related to
the standard progenitor mass function.

\subsection{Relate probability functions to halo mass functions and merger rates}

We define $n(M_d, z_d\vert S_e, \de)dM_d$ as the mean number density of
(descendant) halos of mass between $M_d$ and $M_d+dM_d$ at redshift $z_d$
residing in a region of linear overdensity $\de$ smoothed over scale $S_e$.
This halo mass function is simply related to the conditional probability $P(B|A)$ (denoted as $P_{(B\vert A)}$ below) by
\begin{equation}
\label{N}
n(M_d, z_d\vert S_e, \de)
=\frac{\bar{\rho}}{M_d}\left\vert\frac{dS_d}{dM_d}\right\vert P_{(B|A)}\left(S_d, z_d\vert S_e, \de\right)  \,,
\end{equation}
where $\bar{\rho}$ is the mean mass density.

Similarly, we define $N(M_p, z_p\vert M_d, z_d, S_e, \de)dM_p$ as the mean
number of progenitor halos of mass between $M_p$ and $M_p+dM_p$ at
redshift $z_p$ for a descendant halo of mass $M_d$ and redshift $z_d$
residing in an environment of scale $S_e$ and linear overdensity $\de$.
This progenitor mass function is simply related to the conditional probability $P(C|A,
B)$ (denoted as $P_{(C\vert A, B)}$ below) by
\begin{eqnarray}
\label{Nm}
&&N(M_p, z_p\vert M_d, z_d, S_e, \de)\\ \nonumber
&& \quad  =\frac{M_d}{M_p}\left\vert\frac{dS_p}{dM_p}\right\vert P_{(C|A, B)}\left(S_p, z_p\vert S_d, z_d, S_e, \de\right)  \,.
\end{eqnarray}

The halo merger rate can be written in terms of the progenitor mass
function above.  To this end, we adopt the binary merger assumption as in
\citet{Zhang08}, and define $R(M, \xi, z\vert S_e, \de)$ (same as the $B/n$
term in equation~(8) of \citealt{FM09}) to be the number of mergers per unit
progenitor mass ratio $\xi$ (ratio of the small to the large progenitor
mass) and unit redshift for each descendant halo of mass $M$ at redshift
$z$, under the condition that the linear overdensity on the environmental
scale $S_e$ is $\de$. Due to the binary merger assumption, the merger rate
$R$ can be related to the progenitor mass function via\footnote{Note that
  it is also possible to use $R(M, \xi, z \vert S_e,
  \de)=M (1+\xi)^{-2} dN(M/(1+\xi), z \vert M, z, S_e, \de)/dz\vert_{z'=z}$
  to relate the merger rate to the progenitor mass function. In the limit
  of small $\Delta z$, these two relations should be equivalent. However,
  it has been found that this is generally not true in theories based the
  excursion set. In this paper, we simply use equation~(\ref{rate}),
  which is found to work better in terms of comparison with simulation
  results in \cite{Zhang08}.}
\begin{eqnarray}
\label{rate} 
   && R(M_d, \xi, z_d \vert S_e, \de) \\
    && \quad = \left.\frac{M_d}{(1+\xi)^2}\frac{d}{dz} N\left(\frac{M_d\xi}{1+\xi}, z\vert M_d, z_d, S_e, \de\right)\right\vert_{z=z_d}\,.  \nonumber
\end{eqnarray}

Equations~(\ref{N})-(\ref{rate}) enable us to obtain the
environment-dependent halo mass functions and halo merger rates from
$P(B|A)$ and $P(C|A, B)$. Since $P(B|A) = P(A,B)/P(A)$ and $P(C|A,
B)=P(A,B,C)/P(A,B)$, our next task is therefore to calculate $P(A,B)$ and
$P(A,B,C)$.

\subsection{Express $P(A,B)$ in path integral form}

According to the definition of $P(A, B)$ in \S\ref{Define_rate}, we have
\begin{eqnarray}
\label{PAB1}
&&\int_{S_d}^{\infty}dS_d'P_{AB}(S_d', z_d, S_e, \de)
 = \intnI{\delta_1}\ldots\widehat{d\delta_m}\ldots d\delta_{n} \nonumber \\
&& \qquad \times W(0;\delta_1,\ldots,\delta_m=\delta_e,\ldots,\delta_n;S_d) \,,
\end{eqnarray}
where the positions of $S_e$ and $S_d$ are approximated as $m\epsilon$ and
$n\epsilon$, respectively, with $m$ and $n$ being integers. In other words,
$S_m=S_e$, $S_n=S_d$ and $\delta_m=\delta_e$.  The hat over $d\delta_m$
means that $d\delta_m$ is omitted from the list of integration variables.

By taking partial derivatives with respect to $S_d$ on both sides
of equation~(\ref{PAB1}), and using equation~(\ref{W4}), we obtain
\begin{eqnarray}
\label{PAB2}
&& P(A, B) = P_{AB}(S_d, z_d,\de, S_e)  \\
&& = -\frac{\partial}{\partial S_d}
\intnI{\delta_1}\ldots\widehat{d\delta_m}\ldots d\delta_{n}
\left(1+\frac{1}{2}\sum_{i,j=1}^{n}\Delta_{ij}\partial_i\partial_j\right)\nonumber \\
&& \quad \times W^{gm}(0;\delta_1,\ldots,\delta_m=\delta_e,\ldots,\delta_n;S_d)  \,. \nonumber
\end{eqnarray}
The terms proportional to $\Delta_{ij}$ are the non-Markovian corrections.

We rewrite the summation in the non-Markovian terms in equation~(\ref{PAB2}) as
\begin{equation}
\label{sum_AB}
\frac{1}{2}\sum_{i,j=1}^n\Delta_{ij}\partial_i\partial_j=\sum_{i=1}^{n-1}\Delta_{in}\partial_i\partial_n+\sum_{i<j<n}\Delta_{ij}\partial_i\partial_j  \,,
\end{equation}
where $\Delta_{ii}=0$ for $i=1,2,\ldots,n$ based on
equation~(\ref{Delta_define}) and is therefore not included.  It can be
shown that the first term on the right-hand side of equation~(\ref{sum_AB})
is zero.  The second term can be broken into five pieces, representing all
the possible locations of $i$ and $j$ with respect to $m$ and $n$:
\begin{eqnarray}
\label{sum_AB_break}
\sum_{i<j<n}
&=&\sum_{j=m+1}^{n-1}\cdot\sum_{i=m+1}^{j-1}+\sum_{j=m+1}^{n-1}\cdot(i=m)\\ \nonumber
&+&\sum_{j=m+1}^{n-1}\cdot\sum_{i=1}^{m-1}+(j=m)\cdot\sum_{i=1}^{m-1}+\sum_{j=1}^{m-1}\cdot\sum_{i=1}^{j-1} \,.
\end{eqnarray}
In total, $P(A, B)$ in equation~(\ref{PAB2}) is the sum of the Markovian term and the five
terms in equation~(\ref{sum_AB_break}).  We write these six terms as
\begin{equation}
  P(A, B) = P_{AB}^M + P_{AB}^{NM1} + ... + P_{AB}^{NM5}  \,.
\label{PAB_sum}
\end{equation} 
The superscripts $M$ and $NM$ refer to Markovian and Non-Markovian, respectively,
and the number following each $NM$ refers to the order of the term on the
right-hand side of equation~(\ref{sum_AB_break}).

The algebra involved in deriving these six terms is straightforward but
lengthy.  We leave the details to Appendix A.  The final expression for
$P(A,B)$ is given by equation~(\ref{PAB_final}).

\subsection{Express $P(A,B,C)$ in path integral form}

The derivation of $P(A,B,C)$ is similar to that of $P(A,B)$ above but is
more complicated.  According to the definition of $P(A, B, C)$ in
\S\ref{Define_rate}, we have
\begin{eqnarray}
\label{PABC1}
&&\left(\int_{S_d}^{S_p}dS_d'\int_{S_p}^{\infty}dS_p'
   +\int_{S_p}^{\infty}dS_d'\int_{S_d'}^{\infty}dS_p'\right) \\
&& \qquad  \times P_{ABC}(S_p', z_p, S_d', z_d, S_e, \de)  \nonumber \\
&&=\intnI{\delta_1}\ldots\widehat{d\delta_m}\ldots d\delta_{n}\intnII{\delta_{n+1}}\ldots d\delta_{N}
   \nonumber \\
&& \qquad \times W(0;\delta_1,\ldots,\delta_m=\delta_e,\ldots,\delta_N;S_p)  \,, \nonumber
\end{eqnarray}
where the positions of $S_e$, $S_d$, and $S_p$ are approximated as
$m\epsilon$, $n\epsilon$, and $N\epsilon$, respectively, with $m$, $n$, and
$N$ being integers. In other words, $S_m=S_e$, $S_n=S_d$, $S_N=S_p$, and
$\delta_m=\delta_e$.  The hat over $d\delta_m$ means that $d\delta_m$ is
omitted from the list of integration variables. 

By taking partial derivatives with respect to both $S_p$ and $S_d$ on the
two sides of equation~(\ref{PABC1}), and using equation~(\ref{W4}), we
obtain
\begin{eqnarray}
\label{PABC2}
%
&& P(A, B, C) = P_{ABC}(S_p, z_p, S_d, z_d, S_e, \de)  \\
&& = \partialtwo{S_d}{S_p}\intnI{\delta_1}\ldots\widehat{d\delta_m}\ldots d\delta_{n} 
\intnII {\delta_{n+1}}\ldots d\delta_{N} \nonumber \\
&& 
\left(1+\frac{1}{2}\sum_{i,j=1}^{N}\Delta_{ij}\partial_i\partial_j\right) W^{gm}(0;\delta_1,\ldots,\delta_m=\delta_e,\ldots,\delta_N;S_p)  \,. \nonumber
\end{eqnarray}
The terms proportional to $\Delta_{ij}$ are the non-Markovian corrections.

Similar to equation~(\ref{sum_AB}), we rewrite the summation in the non-Markovian terms above as
\begin{equation}
\label{sum_ABC}
\frac{1}{2}\sum_{i,j=1}^N\Delta_{ij}\partial_i\partial_j=\sum_{i=1}^{N-1}\Delta_{iN}\partial_i\partial_N+\sum_{i<j<N}\Delta_{ij}\partial_i\partial_j  \,.
\end{equation}
As before, the first term here is always zero. We decompose the rest into
thirteen terms:
\begin{eqnarray}
\label{sum_ABC_break}
\sum_{i<j<N} 
&=& \sum_{j=n+1}^{N-1}\cdot\sum_{i=1}^{m-1} + \sum_{j=n+1}^{N-1}\cdot(i=m)\\ \nonumber
&+& \sum_{j=n+1}^{N-1}\cdot\sum_{i=m+1}^{n-1}+ \sum_{j=n+1}^{N-1}\cdot(i=n)\\ \nonumber
&+& \sum_{j=n+1}^{N-1}\cdot\sum_{i=n+1}^{j-1}+ (j=n)\cdot\sum_{i=1}^{m-1}\\ \nonumber
&+& (j=n)\cdot(i=m) + (j=n)\cdot\sum_{i=m+1}^{n-1}\\ \nonumber
&+& \sum_{j=m+1}^{n-1}\cdot\sum_{i=1}^{m-1} + \sum_{j=m+1}^{n-1}\cdot(i=m)\\ \nonumber
&+& \sum_{j=m+1}^{n-1}\cdot\sum_{i=m+1}^{j-1}+ (j=m)\cdot\sum_{i=1}^{m-1}+\sum_{j=1}^{m-1}\cdot\sum_{i=1}^{j-1}  \,.
\end{eqnarray}
Again, we denote the Markovian part of $P(A, B, C)$ as $P_{ABC}^M$, and the
thirteen non-Markovian terms on the right-hand side of
equation~(\ref{sum_ABC_break}) as $P_{ABC}^{NM1}$, ..., and
$P_{ABC}^{NM13}$.  The probability $P(A, B, C)$ is then
\begin{equation} 
  P(A, B, C) = P_{ABC}^M + P_{ABC}^{NM1} + ...
  + P_{ABC}^{NM13} \,.
\label{PABC_sum}
\end{equation}

We leave the details of the derivation of these fourteen terms to Appendix
B.  The final expression for $P(A,B,C)$ is given by
equation~(\ref{PABC_final}).

\subsection{Asymptotic forms in the limit of large environmental scale}
\label{large_scale_limit}

As shown in \S~3.3, 3.4, and Appendix A and B, the general forms of $P(A,
B)$, $P(A, B, C)$, and $P(C\vert A, B)$ contain many terms.  In practice,
it is often unnecessary to consider the general case. Here, we derive the
simplified forms of $P(B|A)$ and $P(C\vert A, B)$ in the limit of large
environmental scale, which is usually the case considered in simulations
and observations.  We leave the details of the derivation to Appendix C and
quote the final results here.  

To linear order in $\delta_e$, the probability of forming a descendant halo
of mass $S_d=S(M_d)$ at redshift $z_d$ that resides in a larger environment
of overdensity $\de$ smoothed over scale $S_e$ is
\begin{eqnarray}
\label{PB_A_approx}
&& P(B|A) = P_{(B|A)}\left(S_d, z_d\vert S_e, \de\right) \\ \nonumber
&\approx&
\frac{\dcI}{ \sqrt{2\pi} S_d^{3/2}}\exp\left(-\frac{\nu^2}{2}\right)\\ \nonumber
&\times&\left\{1-\kappa+\frac{\kappa}{2}\exp\left(\frac{\nu^2}{2}\right)\Gamma\left(0,\frac{\nu^2}{2}\right)\right.\\ \nonumber
&+&\left.\frac{\de}{\dcI}\left[\nu^2-1+\kappa-\frac{\kappa}{2}\exp\left(\frac{\nu^2}{2}\right)\Gamma\left(0,\frac{\nu^2}{2}\right)\right]\right\}  \,,
\end{eqnarray}
where $\nu \equiv \dcI/\sqrt{S_d}$, $\dcI=\delta_c/D(z_d)$ is the barrier
height for forming a descendant halo at redshift $z_d$, $\Gamma(0,x)$ is
the incomplete Gamma function, and $\kappa$ is the non-Markovian parameter
defined in equation~(\ref{Delta_define}).  We note that this equation is
identical to equation~(24) of \citet{Ma11} for the conditional first
crossing rate, which was used to derive the halo bias parameter.  In the
limit of $\delta_e \rightarrow 0$, we recover from
equation~(\ref{PB_A_approx}) the non-Markovian extension of the standard
halo mass function (see, e.g., Table~1 of \citealt{Ma11}):
\begin{eqnarray}
\label{PB}
  P(B) &=& \frac{\dcI}{ \sqrt{2\pi} S_d^{3/2}}\exp\left(-\frac{\nu^2}{2}\right)\\ \nonumber
&\times&\left[1-\kappa+\frac{\kappa}{2}\exp\left(\frac{\nu^2}{2}\right)\Gamma\left(0,\frac{\nu^2}{2}\right)
\right] \,.
\end{eqnarray}

Similarly, the conditional probability (to linear order in $\delta_e$) that
a descendant halo of mass $S_d=S(M_d)$ at redshift $z_d$, residing in a
larger environment of overdensity $\de$ at scale $S_e$, has a progenitor
halo of mass $S_p=S(M_p)$ at redshift $z_p$ (assuming $z_p\approx z_d$) is
\begin{eqnarray}
\label{PC_AB_approx}
&& P(C\vert A,B) 
=  P_{(C|A, B)}\left(S_p, z_p\vert S_d, z_d, S_e, \de\right) \\ \nonumber
&& \approx \frac{\dcII-\dcI}{\sqrt{2\pi}(S_p-S_d)^{3/2}} \left\{1+\kappa 
    \beta\alpha - \sqrt{2\pi}\kappa\nu\left(1-\alpha\right)^{3/2} \right. \\
&& \left. +\pi\kappa\left(\nu^2\frac{\de}{\dcI}-1\right)
  \left(1-\alpha\right)^{3/2}\explarge{\frac{\nu^2}{2}}\erfc{\frac{\nu}{\sqrt{2}}}\right\} \,,  \nonumber
\end{eqnarray}
where $\alpha \equiv S_d/S_p$, and $\beta$ is a simple algebraic function of $\alpha$ 
\begin{equation}
\label{App}
\beta=-2+\frac{(1-\alpha)^{3/2}}{2\alpha}\ln\left( \frac{1+\sqrt{1-\alpha}}{1-\sqrt{1-\alpha}} \right) +\frac{1}{\alpha}+2\alpha \,.
\end{equation}
The variables $\dcII=\delta_c/D(z_p)$ and $\dcI=\delta_c/D(z_d)$ specify
the barrier heights for forming the progenitor and descendant halos at
redshift $z_p$ and $z_d$, respectively.  Equations~(10) and (11) relate
$P_{(C|A, B)}$ above to the mean progenitor mass function $N(M_p, z_p\vert
M_d, z_d, S_e, \de)$ and the merger rate $R(M_d, \xi, z\vert S_e, \de)$.
In the Markovian limit ($\kappa=0$), we note that
equation~(\ref{PC_AB_approx}) reduces to the familiar conditional mass
function of small look-back time ($\dcII-\dcI$) predicted by the excursion
set model, and the dependence on the environmental overdensity $\de$ drops
out. This limit confirms that the introduction of the non-Markovian process
to the excursion set model is the key in introducing the environmental
dependence of halo formation history.

\begin{figure*}
\centering
 \vspace{-0.5in}
   \epsfig{figure=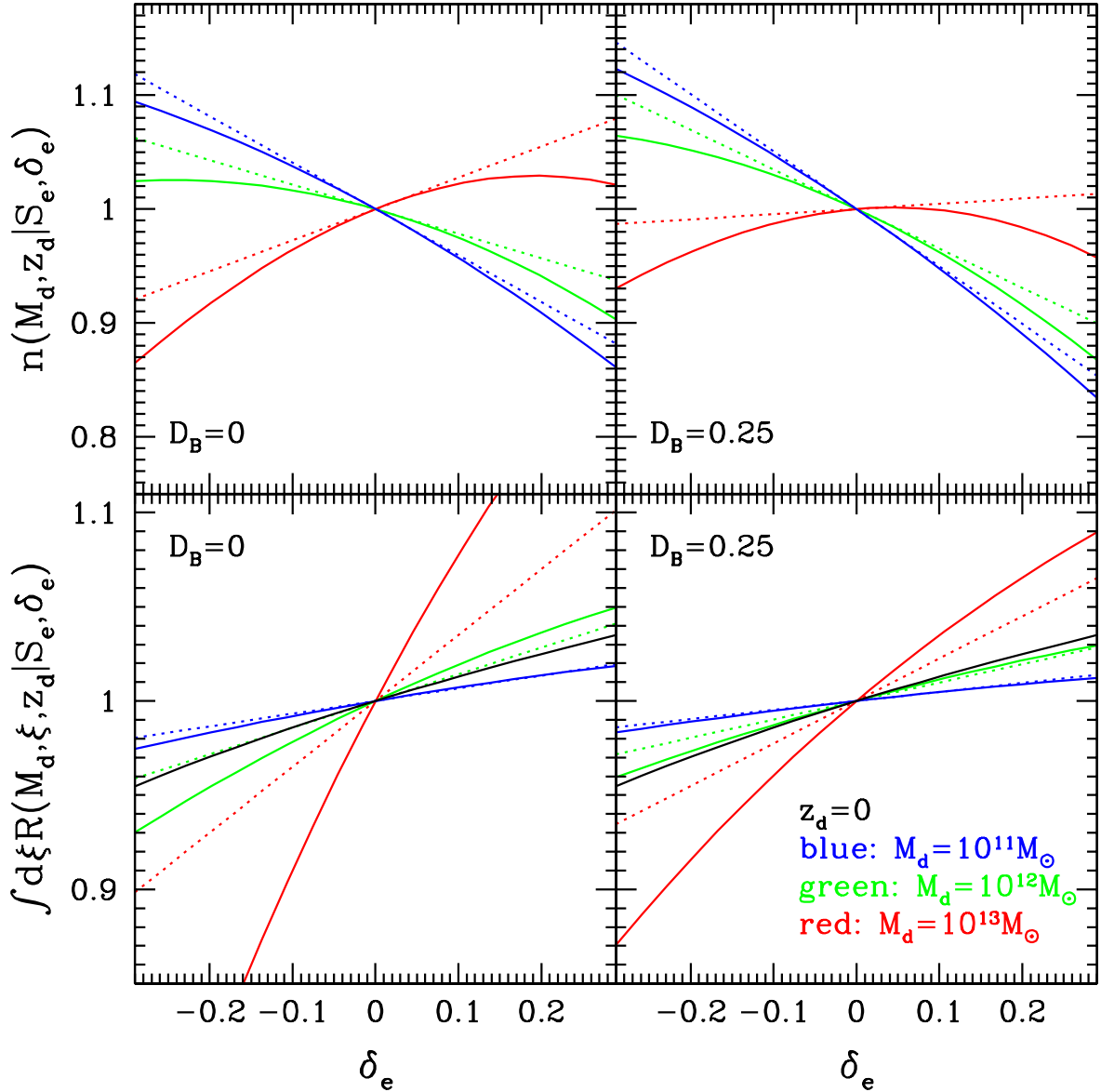,width=6.7in}
   \caption{Environmental dependence of the halo mass function
     $n(M_d,z_d|S_e, \delta_e)$ (upper panels) and the halo merger rate
     $R(M_d,\xi,z_d|S_e,\delta_e)$ for merger mass ratio above 0.01
     (i.e. $\xi=0.01$ to 1) (lower panels).  Two types of barriers in the
     excursion set model are shown for comparison: constant $\delta_c=1.68$
     (left panels) and the diffusing barrier $\delta_c/\sqrt{1+D_B}$ with
     $D_B=0.25$ (right panels).  In each panel, the results are shown for
     three descendant halo masses: $M_d=10^{11}$ (blue), $10^{12}$ (green),
     $10^{13}M_{\odot}$ (red) at redshift $z_d=0$, and the environmental
     mass scale is $S_e=10^{17} M_\odot$.  The vertical axis in each panel is
     normalized to the value when the environmental overdensity field
     $\delta_e$ is zero.  The full expressions from
     equations~(\ref{PAB_final}) and (\ref{PABC_final}) are shown as
     colored solid curves; the approximate expressions (valid to linear
     order in $\delta_e$) from equations~(\ref{PB_A_approx}) and
     (\ref{PC_AB_approx}) are shown as colored dotted curves. As
     equations~(\ref{Nm}) and (\ref{rate}) indicate, the progenitor mass
     function has the same dependence on $\delta_e$ as that in the lower
     panels. The black solid curves in the lower panels plot the merger
     rates from equation~(11) of Fakhouri \& Ma (2009), which is obtained
     from the Millennium simulation.}
\label{lnldensity}
\end{figure*}

Finally, the accuracy of the simple spherical collapse model can be improved by
considering a diffusing barrier instead of a constant one.  The introduction
of the diffusing barrier is motivated by both the elliptical collapse model
and $N$-body studies, for the reason that realistic halos are triaxial rather
than spherical.  For our purpose, we only need to replace $\delta_c$ by
$\delta_c/\sqrt{1+D_B}$ and $\kappa$ by $\kappa/(1+D_B)$ in our formulae to
take into account the diffusing barrier effect \citep{Robertson09, MR2},
with $D_B=0.25$.

\subsection{Numerical results}

In Figure~1, we illustrate the numerical results from our analytic formulae
for the halo mass function $n(M_d,z_d|S_e, \delta_e)$ (upper panels) and
the merger rate $R(M_d,\xi,z_d|S_e,\delta_e)$ (lower panels) as a function
of the halo environment $\delta_e$.
Three descendant halo masses at $z_d=0$ are shown for comparison:
$M_d=10^{11}$ (blue), $10^{12}$ (green), $10^{13}M_{\odot}$ (red).  The
environmental mass scale $S_e$ is chosen to be $10^{17} M_{\odot}$.  The
full expressions (solid curves) are computed from
equations~(\ref{PAB_final}) and (\ref{PABC_final}), and the approximate
expression (dotted curves), valid to linear order in $\delta_e$, are
computed from equations~(\ref{PB_A_approx}) and (\ref{PC_AB_approx}). The
diffusing barrier effect is included in the right two panels with
$D_B=0.25$, and not included in the left two panels (\ie, $D_B=0$). The
cosmological model is a $\Lambda$CDM model with $\Omega_m=0.25$,
$\Omega_b=0.045$, $\Omega_{\Lambda}=0.75$, $h=0.73$, and an initial
power-law power spectrum of the density fluctuation with index $n=1$, and
normalization $\sigma_8=0.9$.

The lower panels of Figure~1 shows a positive dependence of the merger rate
on $\delta_e$.  Since equations~(\ref{Nm}) and (\ref{rate}) indicate that
the progenitor mass function has the same dependence on $\delta_e$ as the
halo merger rate, our results imply that progenitor mass functions are also
higher in regions with higher $\delta_e$.  This environmental trend is
consistent with that seen for halo merger rates in the Millennium
simulation \citep{FM08, FM09, FM10}, where the amplitudes of the merger
rate and progenitor mass functions increase with the environmental
overdensities.  The black solid curves in the lower panels of Figure~1 show
the environmental dependence from the second formula in equation~(11) of
\citet{FM09}.  The larger-scale overdensity in this case is $\delta_7$ and
is measured within a comoving radius of $R=7h^{-1}{\mathrm{Mpc}}$ centered
at each halo in the simulation. As Figure~1 shows, the overall dependence
of the merger rate on $\delta_e$ is similar, while the slope of the curves
from our analytic model has a weak dependence on halo mass.  As discussed
in detail in \citet{FM09}, there are various options for quantifying halo
environment in simulations.  For instance, the environmental overdensity
can be computed by either including or excluding the virial mass of the
central halo within the sphere of radius R over which $\delta_e$ is
computed.  For simplicity, equation~(11) of \citet{FM09} provides two
separate fits for $\delta_7$ and $\delta_{7-{\rm FOF}}$, where the latter
exludes the halo's FOF mass.  They also noted that the difference between
the two definitions, $\delta_7-\delta_{7-{\rm FOF}}$, is a function of halo
mass, increasing from $\sim 0.01$ at $10^{12} M_\odot$ to $\sim 10$ at
$10^{15} M_\odot$.  Given this uncertainty and mass dependence, it is
therefore not surprising that our analytic model predicts mass-dependent
slopes in Figure~1.  A closer comparison between our model prediction and
simulation results would require a more elaborate mapping between the
linear $\delta_e$ in the excursion set model and the nonlinear $\delta_7$
and $\delta_{7-{\rm FOF}}$ used in the simulation. We leave this step to
future studies.

\section{Summary}
\label{discussion}

We have presented a method to introduce ``assembly bias'' into the
excursion set model for the formation and growth of dark matter halos.  Our
calculation is based on the barrier-crossing problem of non-Markovian
processes, which we solve perturbatively using the path integral formalism
developed in MR10.  The new variable that we introduced to parameterize a
halo's larger-scale environment is the linear overdensity field $\delta_e$
smoothed over a chosen scale of $S_e$, where $S_e$ is the variance of the
linear density fluctuations and is a monotonically decreasing function of
the smoothing radius $R$.

To introduce environmental dependence, we isolated $\delta_e$ from the path
integral over the probability density of trajectories $W$ in
equation~(\ref{PAB1}).  We then derived the two main probability functions
$P(A,B)$ and $P(A,B,C)$, defined in Sec~3.1, for forming descendant and
progenitor halos {\it in an environment} in which the linear overdensity
smoothed over scale $S_e$ is given by $\delta_e$.  The calculations are set
up in Sec.~3.3 and 3.4, and the details of how to manipulate the numerous
integrals are given in Appendix A and B.  The final analytic expressions
for $P(A,B)$ and $P(A,B,C)$ are given by equations~(\ref{PAB_final}) and
(\ref{PABC_final}), respectively.

The three key physical quantities that we investigated in this paper are
the descendant halo mass function $n(M_d,z_d|S_e,\delta_e)$, the progenitor
mass function $N(M_p,z_p|M_d,z_d,S_e,\delta_e)$, and the halo merger rate
$R(M_d, \xi, z_d| S_e, d_e)$.  These quantities are related to the
conditional probabilities $P(B|A)$ and $P(C|A,B)$ by
equations~(\ref{N})-(\ref{rate}), which in turn can be computed from our
formulae for $P(A,B)$ and $P(A,B,C)$.

Since the full expressions for the mass functions and merger rates are
complicated, we derived their asymptotic forms in the limit of large
environmental scale (i.e., small $S_e$ and $\delta_e$) in
\S\ref{large_scale_limit} and Appendix C.  This is a useful limit for many
practical purposes.  The approximate expressions for the descendant mass
function and progenitor mass function are given by
equations~(\ref{PB_A_approx}) and (\ref{PC_AB_approx}), respectively.
Figure~1 illustrates the environmental dependence predicted by our model.
It is encouraging that both our analytic calculation and $N$-body results
show that the halo merger rate and progenitor mass function correlate
positively with the environmental density.

The recipe presented in this paper for incorporating environmental
dependence into the excursion set model is quite general. It should provide
a useful theoretical framework for future investigations into how the
spatial distributions and statistical properties of dark matter halos
depend on their mass as well as their assembly history and the larger-scale
environment in which they reside.

\acknowledgments

JZ is supported by the National Science Foundation of China under grant
No. 11273018, the National Basic Research Program of China (2013CB834900),
the national “Thousand Talents Program” for distinguished young scholars, a
grant (No.11DZ2260700) from the Office of Science and Technology in
Shanghai Municipal Government, and the T.D. Lee Scholarship from the Center
for High Energy Physics of Peking University. JZ was previously supported
by the TCC Fellowship of UT Austin and the TAC Fellowship of UC Berkeley,
where a part of this work was done. Support for CPM is provided in part by
grants from the Simons Foundation (\#224959), NASA NNX11AI97G, and
HST-AR-12140.01-A from the Space Telescope Science Institute.  AR is
supported by the Swiss National Science Foundation (SNSF), project ``The
non-Gaussian Universe'' (project number: 200021140236).

\vspace{5 mm}

\section*{\bf Appendix A -- Derivation of $P(A , B)$}

In this appendix, we carry out the integral in equation~(\ref{PAB2}) explicitly
and derive an expression for each of the six terms in the summation in
equation~(\ref{PAB_sum}).  We begin with the following relations from MR10:
\begin{eqnarray}
\label{Wgau_relation}
  && W^{gm}(\delta_0;\delta_1,\ldots,\dn;S_n)
  = W^{gm}(\delta_0;\delta_1,\ldots,\delta_i;S_i)W^{gm}(\delta_i;\delta_{i+1},\ldots,\dn;S_n-S_i)\,,
  \nonumber\\
  && \IIep{\delta_c}{\delta_0}{\dn}{S_n} \equiv \intn{\delta_c}{\delta_1} \ldots
       \intn{\delta_c}{\delta_{n-1}}W^{gm}(\delta_0;\delta_1,\ldots,\delta_n;S_n) \,.
\end{eqnarray}
For the Markovian term in $P(A, B)$ of equation~(\ref{PAB_sum}), we find
\begin{equation}
\label{PAB_M}
  P_{AB}^M=-\partialone{S_n}\intnI{\dn}\IIepI{0}{\de}{S_m} \IIepI{\de}{\dn}{\Snm} \,.
\end{equation}
For the five non-Markovian terms, we find
\begin{eqnarray}
\label{PAB_NM1}
  P_{AB}^{NM1}&=&-\sum_{j=m+1}^{n-1}\sum_{i=m+1}^{j-1}\Delta_{ij}\partialone{S_n}\intnI{\dn}
   \IIepI{0}{\de}{S_m}\IIepI{\de}{\dcI}{\Sim}  \\ \nonumber
&\times&\IIepI{\dcI}{\dcI}{\Sji}\IIepI{\dcI}{\dn}{\Snj} \,, \\ \nonumber
  P_{AB}^{NM2}&=&-\sum_{j=m+1}^{n-1}\Delta_{mj}\partialtwo{\de}{S_n}\intnI{\dn}
   \IIepI{0}{\de}{S_m}\IIepI{\de}{\dcI}{\Sjm}  \IIepI{\dcI}{\dn}{\Snj} \,, \\ \nonumber
  P_{AB}^{NM3}&=&-\sum_{j=m+1}^{n-1}\sum_{i=1}^{m-1}\Delta_{ij}\partialone{S_n}\intnI{\dn}
       \IIepI{0}{\dcI}{S_i}\IIepI{\dcI}{\de}{\Smi}\\ \nonumber
&\times& \IIepI{\de}{\dcI}{\Sjm}\IIepI{\dcI}{\dn}{\Snj} \,, \\ \nonumber
  P_{AB}^{NM4}&=&-\sum_{i=1}^{m-1}\Delta_{im}\partialtwo{\de}{S_n}\intnI{\dn}
       \IIepI{0}{\dcI}{S_i}\IIepI{\dcI}{\de}{\Smi}   \IIepI{\de}{\dn}{\Snm}  \,, \\ \nonumber
  P_{AB}^{NM5}&=&-\sum_{j=1}^{m-1}\sum_{i=1}^{j-1}\Delta_{ij}\partialone{S_n}\intnI{\dn}
    \IIepI{0}{\dcI}{S_i}\IIepI{\dcI}{\dcI}{\Sji}\\ \nonumber
&\times& \IIepI{\dcI}{\de}{\Smj}\IIepI{\de}{\dn}{\Snm} \,.
\end{eqnarray}
To transform the summations into integrations and further simplify these
expressions, we use the following relations from MR10:
\begin{eqnarray}
\label{II_general}
&&\Pi_{\epsilon\rightarrow 0}^{\delta_c}(\delta_0;\delta;S)(\delta_0,\delta \neq \delta_c)
    = \frac{1}{\sqrt{2\pi S}}\left[e^{-(\delta-\delta_0)^2/(2S)}-e^{-(2\delta_c-\delta_0-\delta)^2/(2S)}\right]\,, \quad \Pi_{\epsilon\rightarrow 0}^{\delta_c}(\delta_c;\delta_c;S)
    =\frac{\epsilon}{\sqrt{2\pi}S^{3/2}}  \,,\\ \nonumber
&&\Pi_{\epsilon\rightarrow 0}^{\delta_c}(\delta_0;\delta_c;S)(\delta_0\neq \delta_c)
     =\sqrt{\frac{\epsilon}{\pi}}\frac{\delta_c-\delta_0}{S^{3/2}}e^{-(\delta_c-\delta_0)^2/(2S)}\,,\quad \Pi_{\epsilon\rightarrow 0}^{\delta_c}(\delta_c;\delta;S)(\delta\neq \delta_c)
   = \sqrt{\frac{\epsilon}{\pi}}\frac{\delta_c-\delta}{S^{3/2}}e^{-(\delta_c-\delta)^2/(2S)}\,.
\end{eqnarray}
Substituting these expressions into equations~(\ref{PAB_M}) and (\ref{PAB_NM1}), we obtain
\begin{equation}
\label{iPAB_M}
  P_{AB}^M = -\IIzeroI{0}{\de}{S_m}  \partialone{S_n}\intnI{\dn}\IIzeroI{\de}{\dn}{\Snm} \,,
\end{equation}

\begin{eqnarray}
\label{iPAB_NM1}
  P_{AB}^{NM1} &=&  -\frac{1}{\pi\sqrt{2\pi}}\IIzeroI{0}{\de}{S_m}
   \partialone{S_n}\intnI{\dn}\intS{S_m}{S_n}{S_j}\intS{S_m}{S_j}{S_i}\Deltafunc{i}{j}\\ \nonumber
   &\times&\ratioI{1}{\Sji}\ratioI{\dcIe}{\Sim}\ratioI{\dcIn}{\Snj}
   \explarge{-\ratioII{\dcIe}{\Sim}-\ratioII{\dcIn}{\Snj}}  \,, \\ \nonumber
   P_{AB}^{NM2} &=&  -\frac{1}{\pi}\partialtwo{\de}{S_n}\IIzeroI{0}{\de}{S_m}
   \intnI{\dn}\intS{S_m}{S_n}{S_j}\Deltafunc{m}{j}\\ \nonumber
   &\times&\ratioI{\dcIe}{\Sjm}\ratioI{\dcIn}{\Snj}
   \explarge{-\ratioII{\dcIe}{\Sjm}-\ratioII{\dcIn}{\Snj}} \,, \\ \nonumber
   P_{AB}^{NM3} &=& -\frac{1}{\pi^2} \partialone{S_n}\intnI{\dn}\intS{S_m}{S_n}{S_j}
     \intS{0}{S_m}{S_i}\Deltafunc{i}{j} \frac{\dcI}{S_i^{3/2}}\ratioI{\dcIe}{\Smi}
    \ratioI{\dcIe}{\Sjm}   \\ \nonumber
    &\times & \ratioI{\dcIn}{\Snj}\explarge{-\frac{\dcI^2}{2S_i}}
     \explarge{-\ratioII{\dcIe}{\Smi}-\ratioII{\dcIe}{\Sjm}-\ratioII{\dcIn}{\Snj}} \,, \\ \nonumber
   P_{AB}^{NM4} &=&  -\frac{1}{\pi}\partialone{\de}
   \left[\partialone{S_n}\intnI{\dn}\IIzeroI{\de}{\dn}{\Snm}\right]\\ \nonumber
   &\times &\intS{0}{S_m}{S_i}\Deltafunc{i}{m}\frac{\dcI}{S_i^{3/2}}\ratioI{\dcIe}{\Smi}
   \explarge{-\frac{\dcI^2}{2S_i}-\ratioII{\dcIe}{\Smi}} \,, \\ \nonumber
   P_{AB}^{NM5} &=&  -\frac{1}{\pi\sqrt{2\pi}}
     \left[\partialone{S_n}\intnI{\dn}\IIzeroI{\de}{\dn}{\Snm}\right]\\ \nonumber
&\times &\intS{0}{S_m}{S_j}\intS{0}{S_j}{S_i}\Deltafunc{i}{j}\frac{\dcI}{S_i^{3/2}}\ratioI{\dcIe}{\Sji}
   \ratioI{1}{\Smj}\explarge{-\frac{\dcI^2}{2S_i}-\ratioII{\dcIe}{\Smj}}  \,.
\end{eqnarray}
Using equation~(\ref{Delta_ij}) for $\Deltafunc{i}{j}$, we can work out the
integrals above.  This step is straightforward but tedious, so we only
present the final results here. We also replace $S_m$ and $S_n$ with the
more physical notation for the environment and descendant: $S_m=S_e$ and
$S_n=S_d$.  Our final expression for $P(A, B)$ is given by
\begin{eqnarray}
  \label{PAB_final}
    P(A , B) &=&  P_{AB}^M + P_{AB}^{NM1}+\ldots+P_{AB}^{NM5} \\ \nonumber
    &=& \frac{1}{\sqrt{2\pi}}\ratioI{\dcIe}{\Sde}\left[1+\kappa\frac{S_e}{S_d}\left(1-\frac{(\dcIe)^2}{\Sde}\right)+\kappa\frac{\de(\dcIe)}{S_d}\right]\Eone\IIone\\ \nonumber
    &+&\frac{\kappa}{2\sqrt{2\pi}}(\dcIe)S_d^{-3/2}\Etwo\IIone\Fone+\frac{\kappa}{\pi}\dcI(\dcIe)^2S_e^{-3/2}(\Sde)^{-3/2}\Eone\Ethree\\ \nonumber
    &-&\frac{\kappa}{2}\dcI(\dcIe)S_e^{-1/2}S_d^{-3/2}\Etwo\Ertwo\IIone-\frac{\kappa}{2}\dcI(\dcIe)^2S_e^{-3/2}S_d^{-3/2}\Etwo\Erone\Ertwo\\ \nonumber
    &+&\frac{\kappa}{\sqrt{2\pi}}\dcI(\dcIe)^3S_e^{-1}S_d^{-1}(\Sde)^{-3/2}\Eone\Erone+\frac{\kappa}{\pi}\dcI(\dcIe)^2S_e^{-3/2}(\Sde)^{-3/2}\Eone\Ftwo  \,,
\end{eqnarray}
where
\begin{eqnarray}
\label{definitions}
&&\Eone=\explarge{-\ratioII{\dcIe}{\Sde}} \,, \quad
   \Etwo=\explarge{\frac{(\dcIe)^2}{2S_e}} \,, \quad
    \Ethree=\explarge{-\frac{(2\dcI-\de)^2}{2S_e}}\,,\\ \nonumber
&&\Erone=\erfc{\frac{2\dcI-\de}{\sqrt{2S_e}}} \,, \quad
   \Ertwo=\erfc{\sqrt{\frac{S_d}{2(\Sde)S_e}}(\dcIe)}  \,, \quad
   \Erthree=\erfc{\frac{\dcIe}{\sqrt{2(\Sde)}}}\,,\\ \nonumber
&&\IIone=\IIzeroI{0}{\de}{S_e}\,,\quad {\mathbf F}[a(>0),b]=\int_a^{+\infty}\frac{dx}{x}e^{-(x+b)^2}\,, \\ \nonumber
&&\Fone={\mathbf F}\left[\left(\sqrt{\frac{S_d}{\Sde}}-1\right)\frac{\dcIe}{\sqrt{2S_e}},\frac{\dcIe}{\sqrt{2S_e}}\right] 
   - {\mathbf F}\left[\left(\sqrt{\frac{S_d}{\Sde}}+1\right)\frac{\dcIe}{\sqrt{2S_e}},-\frac{\dcIe}{\sqrt{2S_e}}\right]\,,\\ \nonumber
&&\Ftwo={\mathbf F}\left[\frac{\dcI}{\sqrt{2S_e}},\frac{\dcIe}{\sqrt{2S_e}}\right] \,.
\end{eqnarray}

\vspace{5 mm}

\section*{\bf Appendix B -- Derivation of $P(A , B , C)$}

In this appendix, we carry out the integral in equation~(\ref{PABC2})
explicitly and derive an expression for each of the fourteen terms in the
summation in equation~(\ref{PABC_sum}).  For $P(A, B, C)$ in
equation~(\ref{PABC_sum}), we find the Markovian term to be
\begin{equation}
\label{PABC_M}
  P_{ABC}^M=\partialtwo{S_n}{S_N}\intnI{\dn}\intnII{\dN}
     \IIepI{0}{\de}{S_m}\IIepI{\de}{\dn}{\Snm}  \IIepII{\dn}{\dN}{\SNn} \,,
\end{equation}
and the thirteen non-Markovian terms to be
\begin{eqnarray}
\label{PABC_NM1}
P_{ABC}^{NM1} &=& \sum_{j=n+1}^{N-1}\sum_{i=1}^{m-1}\Delta_{ij}\partialtwo{S_n}{S_N}
 \intnI{\dn}\intnII{\dN} \IIepI{0}{\dcI}{S_i}
  \IIepI{\dcI}{\de}{\Smi}\\ \nonumber
&\times& \IIepI{\de}{\dn}{\Snm}\IIepII{\dn}{\dcII}{\Sjn}\IIepII{\dcII}{\dN}{\SNj} \,,\\ \nonumber
P_{ABC}^{NM2} &=& \sum_{j=n+1}^{N-1}\Delta_{mj}\partialthree{\de}{S_n}{S_N}
    \intnI{\dn}\intnII{\dN}\IIepI{0}{\de}{S_m}
     \IIepI{\de}{\dn}{\Snm}\\ \nonumber
     &\times& \IIepII{\dn}{\dcII}{\Sjn}\IIepII{\dcII}{\dN}{\SNj} \,, \\ \nonumber
P_{ABC}^{NM3} &=& \sum_{j=n+1}^{N-1}\sum_{i=m+1}^{n-1}\Delta_{ij}\partialtwo{S_n}{S_N}
       \intnI{\dn}\intnII{\dN}\IIepI{0}{\de}{S_m}
       \IIepI{\de}{\dcI}{\Sim}\\ \nonumber
     &\times&\IIepI{\dcI}{\dn}{\Sni}\IIepII{\dn}{\dcII}{\Sjn}\IIepII{\dcII}{\dN}{\SNj} \,,\\ \nonumber
 P_{ABC}^{NM4} &=& \sum_{j=n+1}^{N-1}\Delta_{nj}\partialtwo{S_n}{S_N}
          \intnII{\dN}\IIepI{0}{\de}{S_m}\IIepI{\de}{\dcI}{\Snm}\\ \nonumber
         &\times&\IIepII{\dcI}{\dcII}{\Sjn}\IIepII{\dcII}{\dN}{\SNj} \,, \\ \nonumber
 P_{ABC}^{NM5} &=& \sum_{j=n+1}^{N-1}\sum_{i=n+1}^{j-1}\Delta_{ij}\partialtwo{S_n}{S_N}
   \intnI{\dn}\intnII{\dN}\IIepI{0}{\de}{S_m}
    \IIepI{\de}{\dn}{\Snm} \\ \nonumber
&\times&\IIepII{\dn}{\dcII}{\Sin}\IIepII{\dcII}{\dcII}{\Sji}\IIepII{\dcII}{\dN}{\SNj} \,,\\ \nonumber
  P_{ABC}^{NM6}&=&\sum_{i=1}^{m-1}\Delta_{in}\partialtwo{S_n}{S_N}
     \intnII{\dN}\IIepI{0}{\dcI}{S_i}\IIepI{\dcI}{\de}{\Smi}
      \IIepI{\de}{\dcI}{\Snm}\\ \nonumber
&\times&\IIepII{\dcI}{\dN}{\SNn} \,, \\ \nonumber
  P_{ABC}^{NM7}&=&\Delta_{mn}\partialthree{\de}{S_n}{S_N}
   \intnII{\dN}\IIepI{0}{\de}{S_m}\IIepI{\de}{\dcI}{\Snm}
   \IIepII{\dcI}{\dN}{\SNn} \,, \\ \nonumber
  P_{ABC}^{NM8}&=&\sum_{i=m+1}^{n-1}\Delta_{in}\partialtwo{S_n}{S_N}
   \intnII{\dN}\IIepI{0}{\de}{S_m}\IIepI{\de}{\dcI}{\Sim}
    \IIepI{\dcI}{\dcI}{\Sni}\\ \nonumber
&\times&\IIepII{\dcI}{\dN}{\SNn} \,, \\ \nonumber
  P_{ABC}^{NM9} &=& \sum_{j=m+1}^{n-1}\sum_{i=1}^{m-1}\Delta_{ij}\partialtwo{S_n}{S_N}
         \intnI{\dn}\intnII{\dN}\IIepI{0}{\dcI}{S_i}
         \IIepI{\dcI}{\de}{\Smi}\\ \nonumber
        &\times&\IIepI{\de}{\dcI}{\Sjm}\IIepI{\dcI}{\dn}{\Snj}\IIepII{\dn}{\dN}{\SNn} \,,\\ \nonumber
   P_{ABC}^{NM10} &=& \sum_{j=m+1}^{n-1}\Delta_{mj}\partialthree{\de}{S_n}{S_N}
    \intnI{\dn}\intnII{\dN}\IIepI{0}{\de}{S_m}
    \IIepI{\de}{\dcI}{\Sjm}\\ \nonumber
    &\times&\IIepI{\dcI}{\dn}{\Snj}\IIepII{\dn}{\dN}{\SNn} \,, \\ \nonumber
     P_{ABC}^{NM11} &=&  \sum_{j=m+1}^{n-1}\sum_{i=m+1}^{j-1}\Delta_{ij}
     \partialtwo{S_n}{S_N}
     \intnI{\dn}\intnII{\dN}\IIepI{0}{\de}{S_m}
     \IIepI{\de}{\dcI}{\Sim}\\ \nonumber
     &\times&\IIepI{\dcI}{\dcI}{\Sji}\IIepI{\dcI}{\dn}{\Snj}\IIepII{\dn}{\dN}{\SNn} \,,\\ \nonumber
    P_{ABC}^{NM12} &=& \sum_{i=1}^{m-1}\Delta_{im}\partialthree{\de}{S_n}{S_N}
    \intnI{\dn}\intnII{\dN}\IIepI{0}{\dcI}{S_i}
    \IIepI{\dcI}{\de}{\Smi}\\ \nonumber
    &\times&\IIepI{\de}{\dn}{\Snm}\IIepII{\dn}{\dN}{\SNn} \,, \\ \nonumber
   P_{ABC}^{NM13}  &=& \sum_{j=1}^{m-1}\sum_{i=1}^{j-1}\Delta_{ij}\partialtwo{S_n}{S_N}
     \intnI{\dn}\intnII{\dN}\IIepI{0}{\dcI}{S_i}
     \IIepI{\dcI}{\dcI}{\Sji} \\ \nonumber
     &\times&\IIepI{\dcI}{\de}{\Smj}\IIepI{\de}{\dn}{\Snm}\IIepII{\dn}{\dN}{\SNn}  \,.
\end{eqnarray}
The fourteen expressions above can again be written out as
\begin{equation}
\label{iPABC_M}
   P_{ABC}^M=\IIzeroI{0}{\de}{S_m}\partialtwo{S_n}{S_N}\intnI{\dn}
   \intnII{\dN}\IIzeroI{\de}{\dn}{\Snm}\IIzeroII{\dn}{\dN}{\SNn} \,,
\end{equation}

\begin{eqnarray}
\label{iPABC_NM1}
   P_{ABC}^{NM1} &=& \frac{1}{\pi^2}\partialtwo{S_n}{S_N}\intnI{\dn}\intnII{\dN}
   \IIzeroI{\de}{\dn}{\Snm}\intS{0}{S_m}{S_i}\intS{S_n}{S_N}{S_j}\Deltafunc{i}{j}\\ \nonumber
  &\times&\frac{\dcI}{S_i^{3/2}}\ratioI{\dcIe}{\Smi}\ratioI{\dcIIn}{\Sjn}\ratioI{\dcIIN}{\SNj}\explarge{-\frac{\dcI^2}{2S_i}-\ratioII{\dcIe}{\Smi}-\ratioII{\dcIIn}{\Sjn}-\ratioII{\dcIIN}{\SNj}} \,, \\ \nonumber
  P_{ABC}^{NM2} &=& \frac{1}{\pi}\partialthree{\de}{S_n}{S_N}\intnI{\dn}\intnII{\dN}
  \IIzeroI{0}{\de}{S_m}\IIzeroI{\de}{\dn}{\Snm}\\ \nonumber
  &\times&\intS{S_n}{S_N}{S_j}\Deltafunc{m}{j}\ratioI{\dcIIn}{\Sjn}\ratioI{\dcIIN}{\SNj}
  \explarge{-\ratioII{\dcIIn}{\Sjn}-\ratioII{\dcIIN}{\SNj}} \,, \\ \nonumber
  P_{ABC}^{NM3} &=& \frac{1}{\pi^2}\IIzeroI{0}{\de}{S_m}\partialtwo{S_n}{S_N}
     \intnI{\dn}\intnII{\dN}\intS{S_n}{S_N}{S_j}\intS{S_m}{S_n}{S_i}\Deltafunc{i}{j}\\ \nonumber
&\times&\ratioI{\dcIe}{\Sim}\ratioI{\dcIn}{\Sni}\ratioI{\dcIIn}{\Sjn}\ratioI{\dcIIN}{\SNj}\\ \nonumber
&\times&\explarge{-\ratioII{\dcIe}{\Sim}-\ratioII{\dcIn}{\Sni}-\ratioII{\dcIIn}{\Sjn}-\ratioII{\dcIIN}{\SNj}} \,, \\ \nonumber
  P_{ABC}^{NM4}&=&0  \,, \\ \nonumber
 P_{ABC}^{NM5} &=& \frac{1}{\pi\sqrt{2\pi}}\IIzeroI{0}{\de}{S_m}\partialtwo{S_n}{S_N}
       \intnI{\dn}\intnII{\dN}\IIzeroI{\de}{\dn}{\Snm}\\ \nonumber
&\times&\intS{S_n}{S_N}{S_j}\intS{S_n}{S_j}{S_i}\Deltafunc{i}{j}\frac{1}{(\Sji)^{3/2}}
       \ratioI{\dcIIn}{\Sin}\ratioI{\dcIIN}{\SNj}\explarge{-\ratioII{\dcIIn}{\Sin}-\ratioII{\dcIIN}{\SNj}} \,, \\ \nonumber
P_{ABC}^{NM6}&=&P_{ABC}^{NM7}=P_{ABC}^{NM8}=0 \,, \\ \nonumber
  P_{ABC}^{NM9} &=& \frac{1}{\pi^2}\partialtwo{S_n}{S_N}\intnI{\dn}\intnII{\dN} 
      \IIzeroII{\dn}{\dN}{\SNn}\intS{S_m}{S_n}{S_j}\intS{0}{S_m}{S_i}\Deltafunc{i}{j}\\ \nonumber
      &\times& \frac{\dcI}{S_i^{3/2}}\ratioI{\dcIe}{\Smi}\ratioI{\dcIe}{\Sjm}\ratioI{\dcIn}{\Snj}\explarge{-\frac{\dcI^2}{2S_i}-\ratioII{\dcIe}{\Smi}-\ratioII{\dcIe}{\Sjm}-\ratioII{\dcIn}{\Snj}} \,, \\ \nonumber
  P_{ABC}^{NM10} &=& \frac{1}{\pi}\partialthree{\de}{S_n}{S_N}\intnI{\dn}\intnII{\dN}
          \IIzeroI{0}{\de}{S_m}\IIzeroII{\dn}{\dN}{\SNn}\\ \nonumber
          &\times&\intS{S_m}{S_n}{S_j}\Deltafunc{m}{j}\ratioI{\dcIe}{\Sjm}\ratioI{\dcIn}{\Snj}
          \explarge{-\ratioII{\dcIe}{\Sjm}-\ratioII{\dcIn}{\Snj}} \,, \\ \nonumber
   P_{ABC}^{NM11} &=& \frac{1}{\pi\sqrt{2\pi}}\IIzeroI{0}{\de}{S_m}\partialtwo{S_n}{S_N}
   \intnI{\dn}\intnII{\dN}\IIzeroII{\dn}{\dN}{\SNn} \\ \nonumber
   &\times&\intS{S_m}{S_n}{S_j}\intS{S_m}{S_j}{S_i}\Deltafunc{i}{j}
   \frac{1}{(\Sji)^{3/2}}\ratioI{\dcIe}{\Sim}\ratioI{\dcIn}{\Snj}\explarge{-\ratioII{\dcIe}{\Sim}-\ratioII{\dcIn}{\Snj}} \,, \\ \nonumber
   P_{ABC}^{NM12} &=& \frac{1}{\pi}\partialthree{\de}{S_n}{S_N}\intnI{\dn}\intnII{\dN}
   \IIzeroI{\de}{\dn}{\Snm}\IIzeroII{\dn}{\dN}{\SNn}\\ \nonumber
   &\times&\intS{0}{S_m}{S_i}\Deltafunc{i}{m}\frac{\dcI}{S_i^{3/2}}\ratioI{\dcIe}{\Smi}
   \explarge{-\frac{\dcI^2}{2S_i}-\ratioII{\dcIe}{\Smi}} \,, \\ \nonumber
  P_{ABC}^{NM13} &=& \frac{1}{\pi\sqrt{2\pi}}\partialtwo{S_n}{S_N}\intnI{\dn}\intnII{\dN}
  \IIzeroI{\de}{\dn}{\Snm}\IIzeroII{\dn}{\dN}{\SNn}\\ \nonumber
  &\times&\intS{0}{S_m}{S_j}\intS{0}{S_j}{S_i}\Deltafunc{i}{j}
  \frac{1}{(\Sji)^{3/2}}\frac{\dcI}{S_i^{3/2}}\ratioI{\dcIe}{\Smj}
  \explarge{-\frac{\dcI^2}{2S_i}-\ratioII{\dcIe}{\Smj}}  \,.
\end{eqnarray}
For simplicity, the calculation of $P_{ABC}$ is done in the limit of
$\dcII-\dcI\ll 1$.  The lowest order term of the final result is
proportional to $\dcII-\dcI$.  This is because when $\dcII=\dcI$, the
integrals in equation~(\ref{PABC2}) is independent of the
descendent halo mass $S_d$.

We now replace $S_m, S_n$, and $S_N$ with the more physical notation for
the environment, descendant, and progenitor: $S_m=S_e, S_n=S_d$, and
$S_N=S_p$.  Our final expression for $P(A,B,C)$ is
\begin{eqnarray}
\label{PABC_final}
 P(A, B, C)
&=& P_{ABC}^{M} + P_{ABC}^{NM1}+\ldots +P_{ABC}^{NM13}
     \\ \nonumber
&=& \frac{\dcII-\dcI}{2\pi}(\dcIe) (\Sde)^{-3/2}(\Spd)^{-3/2}\Eone\IIone 
    \\ \nonumber
&+& \kappa(\dcII-\dcI)\left\{-\frac{1}{2\sqrt{2\pi}}\dcI(\dcIe)^2S_d^{-3/2}S_p^{-3/2}
   (\Sde)^{-3/2}\Eone\Erone \right.  \\ \nonumber
&+&\frac{1}{2\sqrt{2\pi}}S_e^2S_d^{-3/2}S_p^{-3/2}(\Sde)^{-3/2}\left[1-\frac{\de(\dcIe)}{S_e}+\frac{(\dcIe)^2}{\Sde}\left(1-\frac{2S_d^2}{S_e^2}\right)\right]\Eone\IIone\\ \nonumber
&+&\frac{1}{2\pi}(\dcIe)(\Sde)^{-3/2} (\Spd)^{-3/2}\left[\frac{S_d}{S_p}\mathbf{G_1}+\frac{\de(\dcIe)}{S_d}-\frac{(\dcIe)^2}{\Sde}\frac{S_e}{S_d}\right]\Eone\IIone\\ \nonumber
&+&\frac{1}{2\pi}\dcI(\dcIe)^3S_e^{-1}(\Sde)^{-5/2}(\Spe)^{-2}
         (\Spd)^{-1/2}\Eone\Erone\mathbf{G_2}\\ \nonumber
&-&\frac{1}{2\sqrt{2\pi}}\dcI(\dcIe)^2S_e^{-3/2}S_d^{-3/2} (\Spd)^{-3/2}\Etwo\Erone\Ertwo\\ \nonumber
&-&\frac{1}{2\sqrt{2\pi}}\dcI(\dcIe)S_e^{-1/2}S_d^{-3/2} (\Spd)^{-3/2}\Etwo\Ertwo\IIone\\ \nonumber
&+&\frac{1}{4\pi}(\dcIe)S_d^{-3/2}(\Spd)^{-3/2}\Etwo\IIone\Fone-\frac{1}{2}(\dcIe)S_d^{-3/2}S_p^{-3/2}\Erthree\IIone\\ \nonumber
&+&\frac{1}{2\pi}(\dcIe)S_eS_d^{-3/2}S_p^{-2}(\Sde)^{-1} 
    (\Spd)^{-1/2}\left(S_d-\frac{1}{2}S_p\right)\IIone\\ \nonumber
&\times&\explarge{-\frac{(\dcIe)^2}{2S_e}\left(\frac{S_d}{S_d-S_e}+2\sqrt{\frac{S_d}{S_d-S_e}}\right)}
     \\ \nonumber
&+&\frac{1}{\pi\sqrt{2\pi}}\dcI(\dcIe)^2S_e^{-3/2}(\Sde)^{-3/2}
       \left.(\Spd)^{-3/2}\Eone(\Ethree+\Ftwo)\frac{}{}\right\} \,,
\end{eqnarray}

where
\begin{eqnarray}
\label{A}
\mathbf{G_1} &=& -1 + \frac{1}{2}S_d^{-1}S_p^{-1/2}(\Spd)^{3/2}\ln\frac{\sqrt{S_p}
  + \sqrt{S_p-S_d}}{\sqrt{S_p} - \sqrt{S_p-S_d}} + S_pS_d^{-2}(S_d+S_e)\\ \nonumber
&+&S_d^{-2}S_p^{-1}\left(S_d-\frac{1}{2}S_p\right)\left[2S_d^2+(S_d+S_e)(S_p-S_d)\right]
  - S_d^{-3/2}S_p^{-1}(\Sde)^{1/2}(\Spd)\left(S_d-\frac{1}{2}S_p\right)
\end{eqnarray}
and
\begin{eqnarray}
\label{B} 
\mathbf{G_2}=6S_d-2S_e-4S_p - (\Sde)(\Spd)^{-1}(\Spe) + S_d^{-1}(2S_d-S_e)(\Spe)^2(\Spd)^{-1}  \,.
\end{eqnarray}

\vspace{5 mm}

\section*{\bf Appendix C -- $P(A , B)$ and $P(A , B, C)$ in the limit of large environmental scale}

In the limit of large environmental scale, i.e., small $S_e$, the
overdensity smoothed over this scale, $\de$, also becomes a small parameter
because $\langle\de^2\rangle\sim S_e$.  We will therefore assume $S_e$ is
of the same order as $\de^2$, while keeping in mind that $\delta_e^2/S_e$
is not necessarily small. The conditional probability $P(C\vert A, B)$ is
equal to the ratio of $P(A, B, C)$ in equation~(\ref{PABC_final}) and
$P(A, B)$ in equation~(\ref{PAB_final}), each of which contains special
functions defined in equation~(\ref{definitions}).  The key step in
simplifying $P(C\vert A, B)$ is to find the behavior of these special
functions in the limit of small $S_e$.  After some algebra, we obtain
\begin{eqnarray}
\label{definitions_approx}
&&\Eone\approx\explarge{-\frac{(\dcIe)^2}{2S_d}} \,, \quad
    \IIone\approx\frac{1}{\sqrt{2\pi S_e}}\explarge{-\frac{\de^2}{2S_e}} \,,\quad 
\Erone\approx \frac{\sqrt{2S_e}}{\sqrt{\pi}(2\dcI-\de)}\explarge{-\frac{(2\dcI-\de)^2}{2S_e}} \,, \\ \nonumber
&&\Ertwo\approx \frac{\sqrt{2S_e}}{\sqrt{\pi}(\dcI-\de)}\explarge{-\frac{(\dcI-\de)^2}{2S_e}-\frac{(\dcI-\de)^2}{2S_d}} \,, \quad 
\Erthree\approx\erfc{\frac{\dcIe}{\sqrt{2S_d}}} \,, \\ \nonumber
&&\Fone\approx\explarge{-\frac{(\dcIe)^2}{2S_e}}\Gamma\left[0,\frac{(\dcIe)^2}{2S_d}\right] \,, \quad 
\Ftwo\approx\frac{S_e}{\dcI(2\dcI-\de)}\explarge{-\frac{(2\dcI-\de)^2}{2S_e}} \,.
\end{eqnarray}
Note that $\Etwo$ and $\Ethree$ in equation~(\ref{definitions}) are not
included here, because their forms cannot and need not be further
simplified. The new forms of $\Erone$, $\Ertwo$, and $\Erthree$ are based
on the formula
\begin{equation}
\label{erfc_limit}
\lim_{a\rightarrow +\infty}\erfc{a}\rightarrow\frac{1}{a\sqrt{\pi}}\explarge{-a^2} \,,
\end{equation}
which can be derived from
\begin{eqnarray}
\label{erfc_deriv}
\lim_{a\rightarrow +\infty}\erfc{a}&=&\lim_{a\rightarrow +\infty}\frac{2}{\sqrt{\pi}}\int_a^{+\infty}\exp(-x^2)dx\\ \nonumber
&=&\lim_{a\rightarrow +\infty}\frac{2}{\sqrt{\pi}}\exp(-a^2)\int_a^{+\infty}\exp(a^2-x^2)dx\\ \nonumber
&=&\lim_{a\rightarrow +\infty}\frac{1}{\sqrt{\pi}}\exp(-a^2)\int_0^{+\infty}\frac{\exp(-t)dt}{\sqrt{t+a^2}} \quad\quad \left[{\mathbf Let:}  \quad t=x^2-a^2\right]\\ \nonumber
&=&\lim_{a\rightarrow +\infty}\frac{1}{a\sqrt{\pi}}\exp(-a^2)\int_0^{+\infty}\exp(-t)\left[1+\mathcal{O}\left(\frac{t}{a^2}\right)\right]dt\\ \nonumber
&=&\lim_{a\rightarrow +\infty}\frac{1}{a\sqrt{\pi}}\exp(-a^2)\left[1+\mathcal{O}\left(a^{-2}\right)\right]\\ \nonumber
&\rightarrow&\frac{1}{a\sqrt{\pi}}\explarge{-a^2} \,.
\end{eqnarray}
The simplifications of $\Fone$ and $\Ftwo$ are similar. We need to use the relations
\begin{equation}
\label{F12_limit}
\lim_{a, b\rightarrow +\infty}{\mathbf F}(a,b)\rightarrow\frac{1}{2a(a+b)}\explarge{-(a+b)^2} \,, \quad \lim_{b\rightarrow +\infty, ab\rightarrow c(>0)}\left[{\mathbf F}(a,b)-{\mathbf F}(a+2b,-b)\right]\rightarrow\explarge{-b^2}\Gamma(0,2c)  \,.
\end{equation}
Equation~(\ref{F12_limit}) can be worked out as follows:
\begin{eqnarray}
\label{F2_deriv}
&&\lim_{a, b\rightarrow +\infty}{\mathbf F}[a,b]\\ \nonumber
&=&\lim_{a, b\rightarrow +\infty}\int_a^{+\infty}\explarge{-(x+b)^2}\frac{dx}{x}\\ \nonumber
&=&\lim_{a, b\rightarrow +\infty}\explarge{-(a+b)^2}\int_a^{+\infty}\explarge{(a+b)^2-(x+b)^2}\frac{dx}{x}\\ \nonumber
&=&\lim_{a, b\rightarrow +\infty}\frac{1}{2}\explarge{-(a+b)^2}\int_0^{+\infty}\frac{\exp(-t)dt}{\sqrt{t+(a+b)^2}\left[\sqrt{t+(a+b)^2}-b\right]}  \quad\quad \left[{\mathrm Let:}  \quad t=(x+b)^2-(a+b)^2\right]\\ \nonumber
&=&\lim_{a, b\rightarrow +\infty}\frac{1}{2a(a+b)}\explarge{-(a+b)^2}\int_0^{+\infty}\left\{1+\mathcal{O}\left[\frac{t}{(a+b)^2}\right]+\mathcal{O}\left[\frac{t}{a(a+b)}\right]\right\}\exp(-t)dt\\ \nonumber
&\rightarrow&\frac{1}{2a(a+b)}\explarge{-(a+b)^2} \,,
\end{eqnarray}
\begin{eqnarray}
\label{F1_deriv}
&&\lim_{b\rightarrow +\infty, ab\rightarrow c(>0)}\left[{\mathbf F}(a,b)-{\mathbf F}(a+2b,-b)\right]\\ \nonumber
&=&\lim_{b\rightarrow +\infty, ab\rightarrow c(>0)}\int_a^{+\infty}\frac{dx}{x}\explarge{-(x+b)^2}-\int_{a+2b}^{+\infty}\frac{dx}{x}\explarge{-(x-b)^2}\\ \nonumber
&=&\lim_{b\rightarrow +\infty, ab\rightarrow c(>0)}\int_{a+b}^{+\infty}\frac{dx}{x-b}\exp(-x^2)-\int_{a+b}^{+\infty}\frac{dx}{x+b}\exp(-x^2)\\ \nonumber
&=&\lim_{b\rightarrow +\infty, ab\rightarrow c(>0)}\int_{a+b}^{+\infty}\frac{2b}{x^2-b^2}\exp(-x^2)dx\\ \nonumber
&=&\lim_{b\rightarrow +\infty, ab\rightarrow c(>0)}\explarge{-(a+b)^2} \int_{a+b}^{+\infty}\frac{2b}{x^2-b^2}\explarge{(a+b)^2-x^2}dx\\ \nonumber
&=&\lim_{b\rightarrow +\infty, ab\rightarrow c(>0)}\explarge{-(a+b)^2}b\int_0^{+\infty}\frac{\exp(-t)dt}{\sqrt{t+(a+b)^2}\left[t+(a+b)^2-b^2\right]} \quad\quad \left[{\mathrm Let:}  \quad t=x^2-(a+b)^2\right]\\ \nonumber
&=&\lim_{b\rightarrow +\infty, ab\rightarrow c(>0)}\explarge{-(a+b)^2}\frac{b}{a+b}\int_0^{+\infty}\frac{\exp(-t)dt}{t+2ab+a^2}\left\{1+\mathcal{O}\left[\frac{t}{(a+b)^2}\right]\right\}\\ \nonumber
&\rightarrow&\exp(-b^2-2c)\int_0^{+\infty}\frac{\exp(-t)dt}{t+2c}\\ \nonumber
&\rightarrow&\exp(-b^2)\Gamma(0,2c) \,.
\end{eqnarray}
We are now ready to apply the results of
equation~(\ref{definitions_approx}) to equations~(\ref{PAB_final}) and
(\ref{PABC_final}) for $P(A, B)$ and $P(A,B,C)$, respectively.
Keeping terms up to first order in $\de$ and $\kappa$ as well as terms
proportional to $\de\kappa$, we obtain
\begin{eqnarray}
\label{PAB_approx}
P(A, B)&\approx&\frac{\dcI}{2\pi S_d\sqrt{S_dS_e}}\exp\left(-\frac{\de^2}{2S_e}-\frac{\nu^2}{2}\right)\left\{1-\kappa+\frac{\kappa}{2}\exp\left(\frac{\nu^2}{2}\right)\Gamma\left(0,\frac{\nu^2}{2}\right)\right.\\ \nonumber
&+&\left.\frac{\de}{\dcI}\left[\nu^2-1+\kappa-\frac{\kappa}{2}\exp\left(\frac{\nu^2}{2}\right)\Gamma\left(0,\frac{\nu^2}{2}\right)\right]\right\}  \,,
\end{eqnarray}
\begin{eqnarray}
\label{PABC_approx}
P(A, B, C)&\approx&\frac{(\dcII-\dcI)\dcI}{\left[2\pi S_d(\Spd)\right]^{3/2}\sqrt{S_e}}\exp\left(-\frac{\de^2}{2S_e}-\frac{\nu^2}{2}\right)\\ \nonumber
&\times&\left\{1-\kappa+\beta\alpha\kappa-(1-\alpha)^{3/2}\kappa\left[\sqrt{2\pi}\nu+\pi\exp\left(\frac{\nu^2}{2}\right){\mathrm{erfc}}\left(\frac{\nu}{\sqrt{2}}\right)\right]\right.\\ \nonumber
&+&\frac{\kappa}{2}\exp\left(\frac{\nu^2}{2}\right)\Gamma\left(0,\frac{\nu^2}{2}\right)+\frac{\de}{\dcI}\left[\kappa+(1+\beta\kappa\alpha)(\nu^2-1)+\sqrt{2\pi}\kappa\nu(1-\alpha)^{3/2}(1-\nu^2)\right.\\ \nonumber
&+&\left.\left.\pi\kappa(1-\alpha)^{3/2}\exp\left(\frac{\nu^2}{2}\right){\mathrm{erfc}}\left(\frac{\nu}{\sqrt{2}}\right)-\frac{\kappa}{2}\exp\left(\frac{\nu^2}{2}\right)\Gamma\left(0,\frac{\nu^2}{2}\right)\right]\right\}  \,,
\end{eqnarray}
in which 
\begin{equation}
\label{Ap}
\nu \equiv \frac{\dcI}{\sqrt{S_d}} \,, \qquad \alpha \equiv \frac{S_d}{S_p}\,, \qquad \beta \equiv -2+\frac{(1-\alpha)^{3/2}}{2\alpha}\ln\left( \frac{1+\sqrt{1-\alpha}}{1-\sqrt{1-\alpha}}\right) +\frac{1}{\alpha}+2\alpha   \,.
\end{equation}
Finally, using the results of equations~(\ref{PAB_approx}) and
(\ref{PABC_approx}), we reach the simplified expressions for $P(B|A)$ and
$P(C\vert A, B)$ in equations~(\ref{PB_A_approx}) and (\ref{PC_AB_approx})
of \S\ref{large_scale_limit}.


\label{lastpage}

\end{document}